\documentclass[12pt]{article}
\usepackage{latexsym,amsmath,amssymb,amsbsy,graphicx}
\usepackage[space]{cite} % If exist.

\makeatletter\def\@biblabel#1{\hfill#1.}\makeatother

\allowdisplaybreaks
\begin {document}
% ======================================

\noindent\begin{minipage}{\textwidth}
\begin{center}

{ASTRONOMY, ASTROPHYSICS, AND COSMOLOGY}\\[1cc]

{\Large

ON THE PROBABLE INTERPRETATION OF ANTICORRELATION
BETWEEN THE PROTON TEMPERATURE AND DENSITY IN THE SOLAR WIND

}

\bigskip
\bigskip

{\large Yu.\,V. Dumin$^{1,2}$, A.\,T. Lukashenko$^{3}$,
L.\,M. Svirskaya$^{4,5}$}\\[6pt]

\parbox{.96\textwidth}{\centering\small\it
$^1$M.V.~Lomonosov Moscow State University,
P.K.~Sternberg Astronomical Institute, \\
Universitetskii prosp. 13, Moscow 119234, Russia. \\
E-mail: dumin@sai.msu.ru, dumin@yahoo.com \\
$^2$Space Research Institute of Russian Academy of Sciences, \\
Profsoyuznaya str.\ 84/32, Moscow 117997, Russia. \\
$^3$M.V.~Lomonosov Moscow State University,
D.V.~Skobeltsyn Institute of Nuclear Physics, \\
Leninskie Gory, GSP-1, Moscow 119991, Russia. \\
E-mail: a{\_}lu@mail.ru \\
$^4$South Ural State University (National Research University), \\
Prosp. Lenina 76, Chelyabinsk 454080, Russia. \\
E-mail: svirskaialm@susu.ru, svirskayalm@mail.ru \\
$^5$South Ural State Humanitarian Pedagogical University, \\
Prosp. Lenina 69, Chelyabinsk 454080, Russia.
}\\[1cc]

\smallskip

\end{center}

{\parindent5mm
The anticorrelated distributions of temperature and density of protons
are a well-known property of the solar wind.
Nevertheless, it is unclear till now if they are formed by
some kind of the universal physical mechanism?
Unfortunately, a straightforward comparison of the characteristic
relaxation times for the temperature and density, on the one hand,
and pressure, on the other hand, encounters the problem of
inapplicability of the hydrodynamical approach in the situation when
the free-path length of the protons is considerably greater than
the spatial scale of the structures under consideration.
To resolve this problem, some kinds of the MHD turbulence---reducing
the effective free paths---are usually assumed.
In the present paper, we use an alternative approach based on
the electrostatic (Langmuir) turbulence, described by the mathematical
formalism of the spin-type Hamiltonians, which was actively discussed
in the recent time in the literature on statistical physics.
As follows from the corresponding calculations, formation of
the anticorrelated distributions of temperature and density is
a universal property of the strongly-nonequilibrium plasmas governed by
the spin-type Hamiltonians when they gradually approach the thermodynamic
equilibrium.
So, just this phenomenon could be responsible for the anticorrelations
observed in the solar wind.

\vspace{5pt}\par}

\vspace{1pt}\par

\small UDK: 523.62-726, 533.9. \vspace{1pt}\par
% 523.62-726 Interplanetary plasmas. Solar wind
% 533.9 Physics of plasmas

% 52.25.Kn Thermodynamics of plasmas
% 52.65.-y Plasma simulation
% 94.20.Fg Plasma temperature and density
% 96.50.Ci Solar wind plasma; sources of solar wind
% 96.60.P- Corona
\vspace{1pt}\par
\end{minipage}

% ======================================
\newpage

\section*{Introduction}
\mbox{}\vspace{-\baselineskip}

It is well known that the solar-wind plasma flux, measured by spacecraft,
experiences considerable fluctuations of all basic parameters, such as
the density, temperature, chemical composition, \textit{etc.}
(The general reviews can be found, \textit{e.g.},
in~\cite{Veselovskii_10,Chashei_10} and references therein.)
On the other hand, these fluctuations exhibit the persistent correlations
with each other, which were analysed in detail in a number of previous
works~\cite{Borovsky_12,Veselovsky_18}.

\begin{figure}[t]
\centerline{\includegraphics[width=14cm]{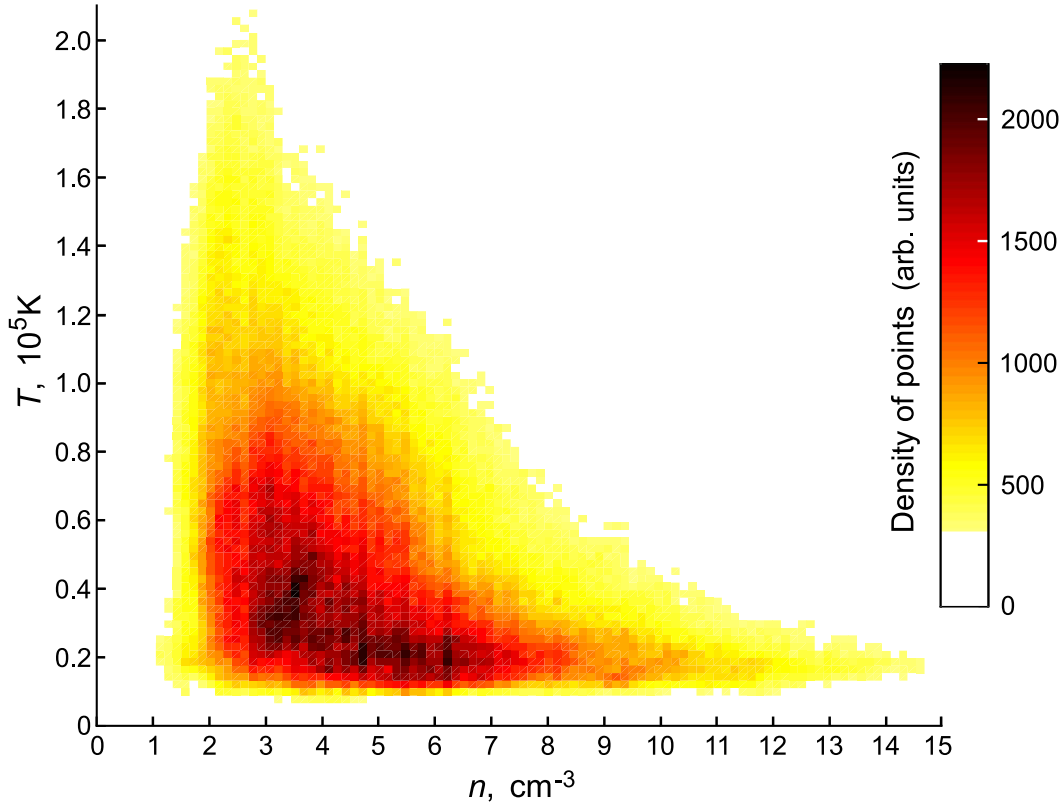}}
\caption{
Correlation diagram of the temperature and density of the solar wind near
the Earth in the period from January 2009 to December 2017.
The density of points is the number of measured values within
a unitary cell of size~$ (0.15\,\mbox{cm}^{-3}){\times}(2200\,\mbox{K}) $.
For better visualization, the values below 300~points per cell were cut off
\label{fig:Corr_Diag}
}
\end{figure}

In the present paper, we pay attention only to one kind of such correlations,
namely, between the temperature of protons (which are the major fraction of
positive ions in the solar wind) and their density.
The corresponding observational data are illustrated in
Fig.~\ref{fig:Corr_Diag}, which is a particular example of the anticorrelated
distributions of~$ T $ and $ n $.
It was drawn by the one-minute values of the respective parameters from
the public database OMNIWeb: \texttt{https://omniweb.gsfc.nasa.gov}\,.
In general, the correlation diagram exhibits a hyperbolic shape, which
corresponds to the negative correlation coefficient.
Here, its particular value is $ r = -0.13 $ for the entire period from
January 2009 to December 2017, \textit{i.e.}, approximately during
a single cycle of the solar activity.
A quite similar value, $ r = -0.15 $, was found in the earlier
work~\cite{Borovsky_12} (see Table~2 in that paper).
Let us emphasize that the above-cited quantities represent the mean values
for the very long time intervals.
The absolute values of the correlation coefficient at the sufficiently short
intervals (about a few hours) can be much greater; \textit{e.g.}, Fig.~2
in paper~\cite{Zank_90}, where $ r = -0.68 $.

One can see in Fig.~\ref{fig:Corr_Diag} the well-expressed horizontal and
vertical tails (parallel to the axes~$ n $ and~$ T $, respectively) in
the distribution of the observational points.
This means that the greater values of density and temperature are
anticorrelated with each other.
Let us mention that we used here the linear rather than logarithmic
scales, which were employed in the most of the previous works.
The above-mentioned tails are well visible just in the linear scale,
while they become compressed into a single spot in the logarithmic scale.
Let us emphasize also that such correlation diagrams can be found also in
a number of other publications by various authors.
The aim of our figure is to demonstrate that the peculiarities found in
the earlier papers survive in the last solar cycle (although a lot of
parameters of the solar wind, in general, changed considerably during
the last two cycles~\cite{Yermolaev_22}).

%%%%%%%%%%%%%%%%%%%%%%%%%%%%%%%%%%%%%%%%%%%%%%%%%%%%%%%%%%%%
%%%%%%%%%%%%%%%%%%%%%%%%%%%%%%%%%%%%%%%%%%%%%%%%%%%%%%%%%%%%

Unfortunately, the most of theoretical works on cosmic plasmas at the present
time are based on the numerical simulations.
They can describe in detail some particular situations but are often unable
to reveal the general physical laws governing the corresponding phenomena.
This situation refers particularly to the interpretation of the above-mentioned
anticorrelations of temperature and density in the solar wind.
Yet another problem is that the theoretical models of the solar wind are
commonly formulated under assumption of incompressibility, just because
the theory of MHD turbulence is developed in detail for the incompressible
medium.
Next, the fluctuations of the density are taken into account by the
perturbation theory, which does not always allow their unambiguous and
self-consistent treatment.
As an example of such an approach, see paper~\cite{Zank_90} and references
therein.

\section{Anticorrelations in the Simple Gas-dynamic Model}
\label{sec:Gas_dyn_model}
\mbox{}\vspace{-\baselineskip}

At the first sight, a rather general interpretation of the anticorrelations
can be derived from the straightforward gas-dynamic analysis of
the characteristic relaxation times for various physical quantities:
\begin{enumerate}
\item
Equilibration of the irregularities in the strongly-nonequilibrium plasmas
should involve both the ``fast'' and ``slow'' stages, as follows from
the two very different time scales for the relaxation of fluctuations in
pressure,~$ ({\Delta}t)_p $, on the one hand, and in the density and
temperature,~$ ({\Delta}t)_n $ and $ ({\Delta}t)_T $, on the other hand.
Indeed, the characteristic rate of the relaxation in pressure is determined
by the speed of sound,%
\footnote{
The term ``relaxation'' does not necessarily imply here a damping of
the oscillations of pressure.
Instead, this could be just an establishment of some average value after
a few oscillations.}
$ c_s\,{\sim}\,\lambda/\tau $ (where $ \lambda $~is the free-path length, and
$ \tau $~is the time of interparticle collisions).
Then, the characteristic time for the relaxation of pressure in the domain of
size~$ \Delta l $ can be estimated as
\begin{equation}
({\Delta}t)_p \approx \Delta l / c_s \sim (\Delta l / \lambda) \, \tau .
\label{eq:Relax_pressure}
\end{equation}
On the other hand, the evolution of density and temperature is described by
the equations of parabolic type:
$ \partial n / \partial t = - {\rm div} (k_n \nabla n) $ and
$ \partial T / \partial t = - {\rm div} (k_T \nabla T) $,
where $ k_{n}\,{\approx}\,k_{T}\,{\sim}\,{\lambda}^2/\tau $.
Consequently, the corresponding relaxation times are estimated as
\begin{equation}
({\Delta}t)_{n,T} \approx (\Delta l)^2 / k_{n,T} \sim
  (\Delta l / \lambda)^2 \tau .
\label{eq:Relax_dens_temp}
\end{equation}
Since $ \Delta l / \lambda \gg 1 $, then
$ ({\Delta}t)_{n,T} \gg ({\Delta}t)_p $, \textit{i.e.}, the equilibrium value
of pressure throughout the entire system should be established very quickly,
while the fluctuations of density and temperature relax at the much longer
time interval.
\item
In the conditions when a mean pressure was established, it follows immediately
from the equation of state for an ideal gas, $ p\,{=}\,nkT $ (where $ k $~is
the Boltzmann constant), that the long-lived fluctuations of
density~$ {\Delta}n $ and temperature~$ {\Delta}T $ will be anticorrelated
with each other: the smaller is the density, the greater should be
the temperature, and \textit{visa versa}.
\end{enumerate}

Unfortunately, despite of universality of the above arguments, their
application to the case of the solar wind encounters the problem of
applicability of the hydrodynamic approximation.
Really, if we take the typical proton temperature $ T \approx 10^5$\,K and
their typical density $ n \approx 10$\,cm${}^{-3}$~\cite{Chashei_10} and
then use the well-known expressions for scattering of charged particles
by each other (\textit{e.g.}, formulas~(3.14) and (3.15) in~\cite{Lang_74}),
we get the free-path length%
\footnote{
We imply here the scattering of protons by protons, while the electrons
represent a uniform neutralizing background.}
$ {\lambda}{\approx}10^9$\,km.
This value is very large even in comparison with the total distance from
the Sun to the Earth, $ 1.5{\cdot}10^8$\,km.
On the other hand, the above-discussed anticorrelations correspond to
the characteristic scales of irregularities that are less by a few orders of
magnitude.
Indeed, since the typical speed of the solar wind passing through
the measuring apparatus is about 500\,km/s, then the fluctuations at
the time scale of a few minutes correspond to the spatial fluctuations
of size about a few tens or hundreds of thousands of kilometers.
Therefore, to justify the applicability of the hydrodynamic approximation,
the commonly-used models of the solar wind (see, for example,~\cite{Chashei_97}
and references therein) assume a presence of some kind of the MHD turbulence,
considerably increasing the effective collisional frequencies.

It is the aim of the present paper to demonstrate that yet another approach
to solving the above-mentioned problem might be the introduction of
the electrostatic (Langmuir) rather than magnetohydrodynamic turbulence.
A convenient mathematical formalism for the description of such turbulence
are models of the nonequilibrium plasmas based on the spin-type Hamiltonians,
which are actively developed in the last years.

Finally, let us mention that the anticorrelations of temperature and density
in the framework of hydrodynamic equations were predicted already in
the very old paper~\cite{Zank_90}; see equation~(10) from that article.
Unfortunately, this result depended crucially of the employed version of
the perturbation theory: one variant of perturbations (which was called by
the authors ``Type~I'') really resulted in the anticorrelations, while
the second variant (``Type~II'') led to the positive correlations.
So, it remained unclear which of these cases is realized in the particular
physical situations.

\section{Anticorrelations in the Strongly-Nonequilibrium \\
Plasmas Described by the Spin-Type Hamiltonians}
\label{sec:Spin_Hamilt}
\mbox{}\vspace{-\baselineskip}

The method of description of the long-range (particularly, Coulomb) many-body
systems by the spin-type Hamiltonians originated in the literature on
statistical physics about 10~years ago (see, for
example,~\cite{Casetti_14,Teles_15} and references therein)
but remains poorly known in the plasma-physics community till now.
Therefore, it is reasonable to resume here its basic ideas.

As is known, the Hamiltonian theory of the spin systems is a very old and
mathematically well-developed branch of theoretical physics, whose history
lasts for almost a century~\cite{Isihara_71,Rumer_72}.
Its original and most important scope of applicability was the solid-state
physics and, first of all, the theory of magnetic phenomena.
However, the spin-type Hamiltonians were used subsequently also for modelling
a lot of other physical phenomena, \textit{e.g.}, the strongly-nonequilibrium
phase transformations in various theories with the spontaneous symmetry
breaking~\cite{Dumin_09,Dumin_14}, and even in the adjacent scientific
disciplines, such as biophysics.

\begin{figure}[t]
\begin{center}
\includegraphics[width=0.85\textwidth]{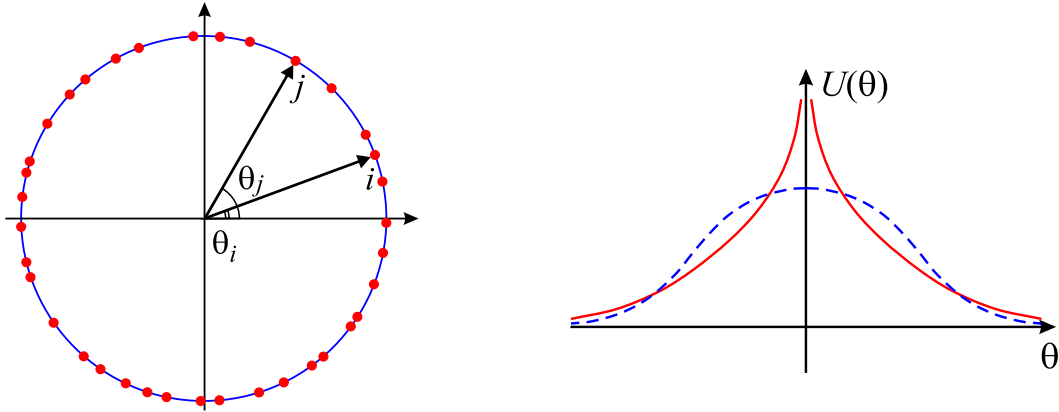}
\end{center}
\caption{
Description of plasmas by the spin-type models:
(left-hand panel) geometric configuration of the one-dimensional system, and
(right-hand panel) approximation of the Coulomb potential of interparticle
interaction (solid red curve) by the Fourier harmonic of the lowest order
(dashed blue curve)
\label{fig:Model}
}
\end{figure}

As regards plasmas and the ensembles of particles with gravitational
(Newtonian) interaction, the spin-type Hamiltonians are introduced there
in the following way.
Let us consider a one-dimensional system of ions, which is a reasonable
approximation of the strongly-magnetized plasmas.
Since the plasma as a whole should be electrically neutral,
the set of ions is assumed to be immersed into a uniform electronic gas,
which corresponds to the well-known approximation of the ``one-component
plasma'' (OCP)~\cite{Ichimaru_82}.
Next, for the sake of simplicity we shall assume the periodic boundary
conditions; so that this system will be topologically equivalent to
the ring.
Then, from the mathematical point of view, positions of the ions can be
formally characterized by the angles~$ {\theta}_i $, as shown in the left
panel of Fig.~\ref{fig:Model}.

%%%%%%%%%%%%%%%%%%%%%%%%%%%%%%%%%%%%%
%%%%%%%%%%%%%%%%%%%%%%%%%%%%%%%%%%%%%

Next, the Coulomb potential~$ U(\theta) $ can be approximated in the simplest
case by a single Fourier harmonic of the lowest order, \textit{i.e.},
the cosine function with period corresponding to the size of the entire system,
as shown in the right panel of the same figure.
Strictly speaking, the Fourier expansion of the Coulomb potential is ill
defined because of its singularity in zero.
So, the approximation by only one (the first) harmonic is a quite crude
approximation of the general shape of potential.
Nevertheless, since the first harmonic ``globally'' covers the entire system,
it can reasonably represent the long-range character of the Coulomb (or
gravitational) interactions.

Finally, Hamiltonian of the ions can be written in the form similar to
the so-called Heisenberg Hamiltonian for a spin system:
\begin{equation}
{\cal H} = \, \frac{1}{2} \sum\limits_{i=1}^N p_i^2 +
           \frac{J}{N} \sum\limits_{i=1}^N \sum\limits_{j=1}^{i-1}
           \big[ 1 - \cos({\theta}_i - {\theta}_j)\big] \, ,
\label{eq:Hamiltonian}
\end{equation}
where
$ p_i $~are the momenta of ions (which are assumed for simplicity to be of
the unitary masses),
$ N $~is their total number, and
$ J $~is the coupling parameter commonly used in the spin models, which
can be expressed through the parameters of the Coulomb or gravitational system.
Let us mention that in the case of plasmas, where particles of the same charge
are repelled from each other, the coupling parameter~$ J $ in
formula~(\ref{eq:Hamiltonian}) should be negative.
Following the terminology used in the solid-state physics, this corresponds to
the ``antiferromagnetic'' type of interactions.
On the other hand, at positive~$ J $ (``ferromagnetic'' interaction)
the particles are attracted to each other; this corresponds to the gravitating
systems.

In summary, the interaction between spins by the cosine law is analogous to
the Coulomb interaction between the charged particles; the angle between
two spins plying the role of distance between the particles in plasma.
(The additional constant term in this formula was introduced just to get
an exact equivalence with the spin system; it evidently does not affect
the equations of motion.)
Let us emphasize that the interaction between the ions is ``collective'':
it is irreducible to a sequence of pair collisions.
Besides, the Hamiltonian~(\ref{eq:Hamiltonian}) evidently takes into account
only electrical, not magnetic forces.
Therefore, it describes the processes similar to the electrostatic (Langmuir)
waves in plasmas.

%%%%%%%%%%%%%%%%%%%%%%%%%%%%%%%%%%%%%%%%%%%%%%%%%%%%%%%%%%%%
\begin{figure}[t]
\centerline{\includegraphics[width=14cm]{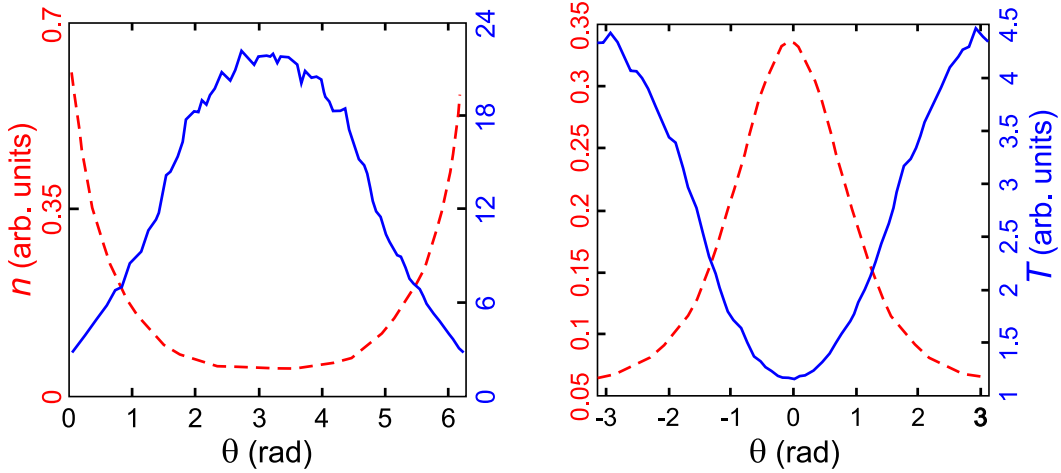}}
\caption{
Examples of the anticorrelated distributions of temperature (solid blue
curves) and density (dashed red curves), computed for the systems described
by the spin-type Hamiltonians: (left-hand paned) Fig.~2 from
paper~\cite{Casetti_14}, reprinted with permission from Springer Nature
Switzerland AG, {\copyright}~2014; and (right-hand panel) Fig.~2 from
paper~\cite{Teles_15}, reprinted with permission from American Physical
Society, {\copyright}~2015. The results are plotted in both panels in
the dimensionless units used in the original works
\label{fig:Simulations}
}
\end{figure}
%%%%%%%%%%%%%%%%%%%%%%%%%%%%%%%%%%%%%%%%%%%%%%%%%%%%%%%%%%%%

As is known, the spin systems are characterized by the universal thermodynamic
properties; and their equilibrium parameters can be usually calculated
analytically, at least, in the one- and two-dimensional
cases~\cite{Isihara_71,Rumer_72}.
On the other hand, the nonequilibrium properties depend on the initial and/or
boundary conditions; and they are usually derived by combining the analytical
and numerical methods.
Particularly, dynamics of the strongly-nonequilibrium plasmas described by
the Hamiltonian~(\ref{eq:Hamiltonian}) was studied in detail in
papers~\cite{Casetti_14,Teles_15}.
Examples of these calculations are shown in Fig.~\ref{fig:Simulations}.
The left panel refers to the case of ``antiferromagnetic'' (Coulomb)
interaction, when an external constant field was present, and the initial
velocity distribution was non-Maxwellian.
The right panel refers to the case of ``ferromagnetic'' (gravitational)
interaction, with the Maxwellian initial velocity distribution perturbed by
a short pulse of the external field.

%%%%%%%%%%%%%%%%%%%%%%%%%%%%%%%%%%%%%%%%%%%%%%%%%%%%%%%%%%%%
\begin{figure}[t]
\centerline{\includegraphics[width=16cm]{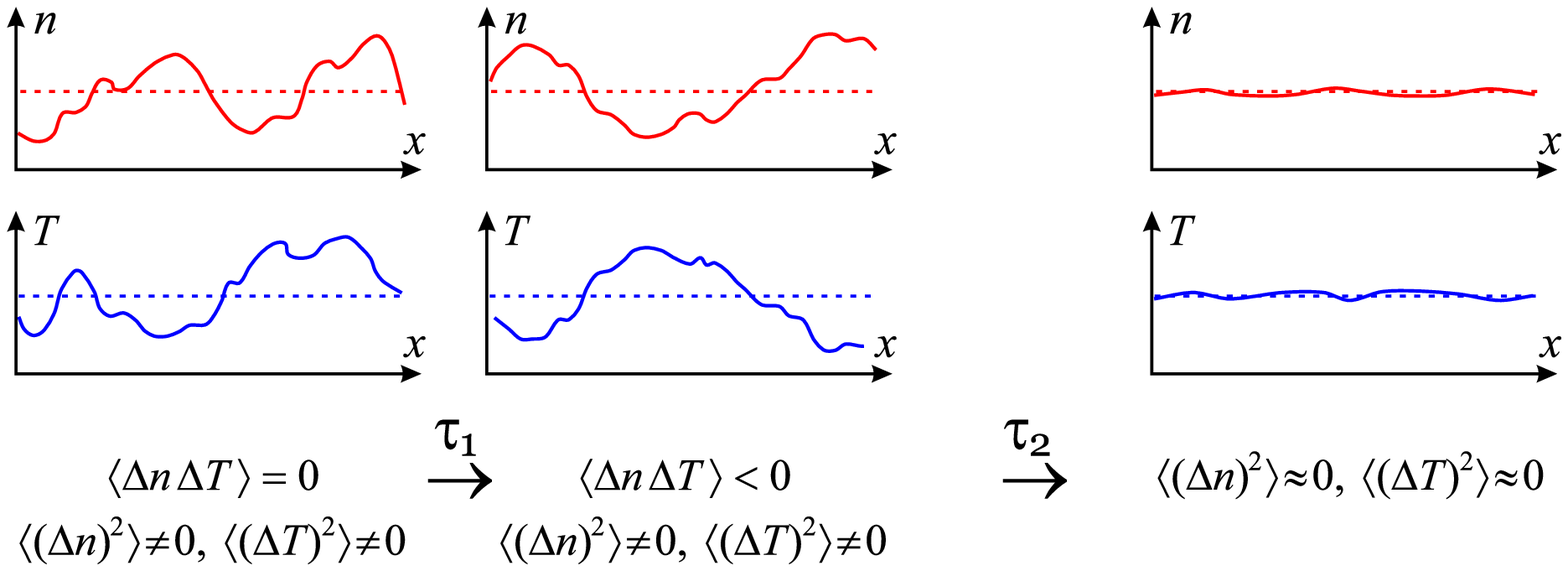}}
\caption{
Sketch of two stages in the relaxation of strongly-nonequilibrium plasmas.
The enhanced space between the second and third panels implies a much greater
duration of the second stage of the relaxation as compared to the first stage
\label{fig:Relaxation}
}
\end{figure}
%%%%%%%%%%%%%%%%%%%%%%%%%%%%%%%%%%%%%%%%%%%%%%%%%%%%%%%%%%%%

Therefore, it is seen from these plots that a generic property of
the relaxation of the nonequilibrium many-body systems described by the
spin-type Hamiltonians is a formation of the anticorrelated spatial
distributions of temperature and density at some intermediate stage of
their evolution, before approaching the complete thermodynamic equilibrium.
The entire such process can be illustrated by the diagram shown in
Fig.~\ref{fig:Relaxation}:
\begin{enumerate}
\item
It is assumed that the plasma was initially created in the state with
considerable fluctuations of both the temperature~$ T $ and density~$ n $
that were independent of each other (left panel).
\item
Next, this plasma quickly evolves to the quasi-stationary state where
large fluctuations~$ \Delta T $ and $ \Delta n $ still survive but become
anticorrelated with each other (middle panel).
\item
Finally, at the much longer time interval, these fluctuations gradually
decay, and the complete thermodynamic equilibrium is established (right
panel).
\end{enumerate}

It is important to emphasize that the characteristic time scales of the first
and second stages of the relaxation, $ {\tau}_1 $ and $ {\tau}_2 $, are
radically different from each other~\cite{Casetti_14,Teles_15}.
The first stage (\textit{i.e.}, transition from the arbitrary nonuniform state
to the long-lived quasi-stationary state with anticorrelated distributions
of~$ T $ and~$ n $) takes about one unit of the dimensionless time, if
the Hamiltonian~(\ref{eq:Hamiltonian}) was written in the dimensionless form.
In the physical units, this corresponds to the inverse plasma frequency:
\begin{equation}
{\tau}_1 \sim 1 / {\nu}_{\rm pl} \, .
\label{eq:tau_1}
\end{equation}
On the other hand, the second state (\textit{i.e.}, a relaxation of
the long-lived quasi-stationary state to the complete thermodynamic
equilibrium) takes the time:
\begin{equation}
{\tau}_2 \sim  N / {\nu}_{\rm pl} \, ,
\label{eq:tau_2}
\end{equation}
where $ N $~is the total number of the charged particles participating in this
process.
Since $ N \gg 1 $, then $ {\tau}_2 \gg {\tau}_1 $.

Just the long-lived ``intermediate'' state is of considerable interest for
the interpretation of a few observable physical phenomena.
For example, it was suggested already in the original paper~\cite{Teles_15}
that the intermediate anticorrelated state of~$ T $ and $ n $ might be employed
for solving the well-known astrophysical problem of heating the solar corona,
namely, a sharp increase of its temperature from~$ {\sim}5000 $\,K at the level
of photosphere to~$ (1-2){\cdot}10^6 $\,K in the
corona~\cite{Walsh_03,Erdelyi_07}.

Unfortunately, a more careful analysis shows that the above-mentioned mechanism
cannot be related to heating of the solar corona for the following two
reasons~\cite{Dumin_16}.
First, it assumes that the irregularities (large-scale fluctuations) of
temperature and density are established in the course of self-consistent
relaxation in the Coulomb system (plasma) under consideration, whereas a sharp
decay of the density in the solar corona is evidently associated with
gravitational stratification in the external field.
Second, even if the large-scale spatial variations of temperature in
the solar corona were formed by the process of relaxation predicted by
the spin-type model, they would be directed randomly (upwards and downwards)
in the various magnetic-flux tubes.
As a result, the observed temperature would either change in the opposite
directions in various regions along the solar surface or remain approximately
constant (if the individual magnetic tubes were sufficiently thin and
unresolved by the observations).
Both these cases are evidently irrelevant to the actual situation.

On the other hand, the two-step relaxation with a long-lived intermediate
anticorrelated stage, as depicted in Fig.~\ref{fig:Relaxation}, is of
considerable interest just for the interpretation of the irregular structure
of the solar wind.
Indeed, both the above-mentioned obstacles will no longer take place here:
\begin{enumerate}
\item
Since the solar wind propagates more or less freely, the gravitational field
no longer plays a significant role in the local reference frame associated with
the substance (\textit{i.e.}, the situation becomes in a certain sense similar to
the state of zero gravity inside a spaceship).
Therefore, the relaxation of the temperature and density can proceed
approximately in a self-consistent manner, as required in the spin-type plasma
model.
\item
The distributions of the solar-wind temperature and density in the comoving
reference frame really exhibit the {\it oppositely-directed}\/ fluctuations
of various spatial scales.
(When the solar wind passes through the measuring apparatus onboard
a spacecraft, these fluctuations are recorded as the changes in time.)
\end{enumerate}

Therefore, Fig.~\ref{fig:Corr_Diag} and the results of other papers cited in
the Introduction fit well with the general predictions of the spin-type plasma
models.
Namely, the uncorrelated fluctuations of~$ T $ and $ n $, naturally expected
in outflow of the solar wind from the corona (left panel in
Fig.~\ref{fig:Relaxation}), are transformed to the long-lived anticorrelated
distributions of these quantities (middle panel in the same figure).

Let us estimate the characteristic times of the respective processes by
formulas~(\ref{eq:tau_1}) and~(\ref{eq:tau_2}).
For the typical concentration of protons in the solar wind
$ n \approx 10$\,cm${}^{-3}$, we find that the anticorrelations are formed
at the time scale
$ {\tau}_1 \sim 10^{-3}$\,s, which is evidently much less than the one-minute
scale of the fluctuations discussed above.

Next, to evaluate the lifetime of the anticorrelated state~$ {\tau}_2 $,
we need to know the effective number of particles~$ N $ in the
quasi-one-dimensional ``spin chain''.
It can be estimated as~$ n^{1/3} {\Delta}l $, where
$ n^{1/3} $~is the mean interparticle separation, and
$ {\Delta}l $~is the characteristic spatial scale of the irregularity.
At the typical speed of the solar wind~500\,km/s, we get for the
above-mentioned one-minute fluctuations:
$ {\Delta}l \approx 3{\cdot}10^9$\,cm.
Then, formula~(\ref{eq:tau_2}) leads to $ {\tau}_2 \sim 5{\cdot}10^{6}$\,s,
\textit{i.e.}, the anticorrelations can well survive even at the very long
time intervals.
In a sense, this estimate even looks suspiciously optimistic: it is possible
that the effects of MHD turbulence destroy the anticorrelated states at
the shorter time scales.
This problem still needs to be studied in more detail.

\section*{Discussion and Conclusions}
\mbox{}\vspace{-\baselineskip}

\begin{figure}[t]
\centerline{\includegraphics[width=10.5cm]{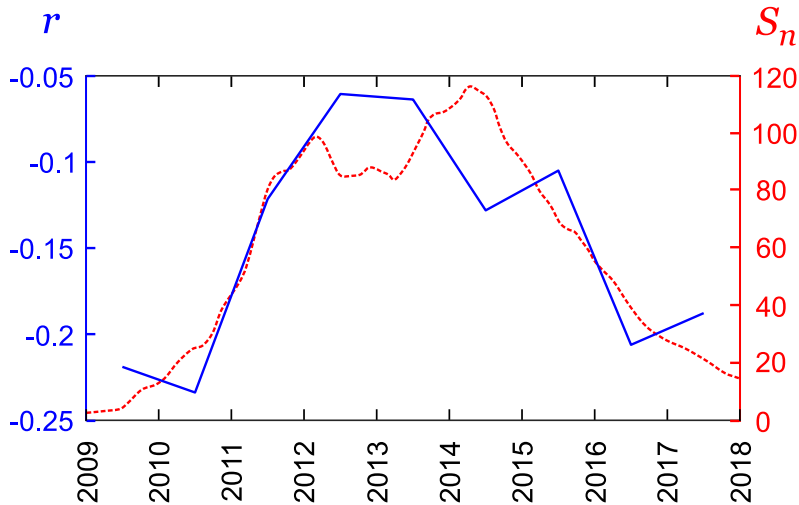}}
\caption{
Variations of the correlation coefficient~$ r $ (solid blue curve)
depending on the level of solar activity, characterized by the
mean monthly sunspot number~$ S_n $ (dotted red curve) during
the whole solar cycle, from 2009 to 2018.
\label{fig:Corr_Sunspot_Time}
}
\end{figure}

It is interesting to discuss in more detail the efficiency of formation
of the anticorrelated distributions of temperature and density in the various
phases of the solar activity cycle.
Indeed, the coefficient of correlation $ r = -0.13 $ mentioned in the
Introduction (and similar values found by other authors) was obtained for
a sufficiently long period, from January 2009 to December 2017 inclusively,
\textit{i.e.}, approximately for a whole cycle.
On the other hand, as follows from the detailed processing of the observational
data, the coefficient~$ r $ can vary substantially depending on the level of
solar activity at the shorter time intervals.
For example, as seen in Fig.~\ref{fig:Corr_Sunspot_Time}, the absolute value of
the correlation coefficient~$ | r | $ reaches the maximum values
about~0.2--0.25 in the periods of low solar activity (2009--2010 and
2016--2018) and drops down to~0.05 in the period of maximum activity.

%%%%%%%%%%%%%%%%%%%%%%%%%%%%%%%%%%%%%%%%%%%%%%%%%%%%%%%%%%%%
%%%%%%%%%%%%%%%%%%%%%%%%%%%%%%%%%%%%%%%%%%%%%%%%%%%%%%%%%%%%

In general, such a behaviour is not surprising: the outflow of the solar wind
from the corona in the periods of low activity should have a more ``gentle''
character, when fluctuations of temperature and density are initially random
and independent from each other.
Just this assumption is the basis of the model of plasma relaxation described
by the spin-type Hamiltonians.
On the other hand, the periods of high solar activity are characterized
by a larger number of the events where the initial fluctuations of~$ T $ and
$ n $ are correlated positively, \textit{e.g.}, due to the coronal mass
ejections (CME) caused by a disruption of the magnetic tubes in the solar
atmosphere.
Indeed, as seen in Fig.~2 from paper~\cite{Yermolaev_22}, the types of
the solar wind associated with ejections take up a significantly larger
percentage of time in the periods of high solar activity than in the periods
of low activity.
The initial positive correlations between~$ T $ and $ n $ in the region of
solar wind formation should partially survive during its subsequent evolution
in propagation to the Earth, thereby reducing the magnitude of anticorrelations
in the respective periods of observations.

Finally, to avoid misunderstanding, let us emphasize that the degree of
universality of the above mechanism of anticorrelations should not be
exaggerated.
For example, the anticorrelated distributions of temperature and density
are well known in the planetary atmospheres~\cite{Rishbeth_69,Ratcliffe_72},
but they have there a completely different nature.
Roughly speaking, the increase of temperature in the rarefied outer layers
of the planetary atmospheres is caused by the fact that they are located
closer to the external source of irradiation---the Sun---and, therefore,
absorb its radiation more efficiently.
This is unrelated to the processes of relaxation in a non-equilibrium gas.

In summary, we can conclude that the method of the spin-type Hamiltonians
is an efficient way to describe Langmuir turbulence in the
strongly-nonequilibrium plasmas.
On the one hand, this approach is based on a number of simplifying assumptions,
particularly, the model of ``one-component'' plasma and a quite rude
approximation of the Coulomb potential, as discussed above.
But, on the other hand, this simplification enables us to construct
a self-consistent model for relaxation of the plasma irregularities, without
any additional assumptions about the characteristics of the resulting
turbulence.
The two-step relaxation of the strongly non-equilibrium plasmas with
a long-lived intermediate stage of anticorrelated distributions of temperature
and density, predicted in the framework of this model, is of considerable
interest for the interpretation of the observed properties of the solar wind.

\bigskip

YVD is grateful to S.\,I.~Bezrodnykh for valuable advices on the mathematical
issues, as well as to B.\,P.~Filippov, I.\,F.~Nikulin, M.\,O.~Riazantseva,
and Yu.\,I.~Yermolaev for the discussion of observational data.
Besides, we are deeply grateful to the anonymous referee for a number of
interesting comments and suggestions, considerably improved the manuscript.
The solar wind data were obtained from the GSFC/SPDF OMNIWeb interface at
\texttt{https://omniweb.gsfc.nasa.gov}\,.

% ======================================

% ======================================
\end {document}